# Volume-Independent Music Matching by Frequency Spectrum Comparison

Anthony Lee

Milton Academy

March 2, 2022

# Table of Contents

# Abstract

Often, I hear a piece of music and wonder what the name of the piece is. Indeed, there are applications such as Shazam app that provides music matching. However, the limitations of those apps are that the same piece performed by the same musician cannot be identified if it is not the same recording. Shazam identifies the recording of it, not the music. This is because Shazam matches the variation in volume, not the frequencies of the sound. This research attempts to match music the way humans understand it: by the frequency spectrum of music, not the volume variation.

Essentially, the idea is to pre-compute the frequency spectrums of all the music in the database, then take the unknown piece and try to match its frequency spectrum against every segment of every music in the database. I did it with by matching the frequency spectrum of the unknown piece to our database by sliding the window by 0.1 seconds and calculating the error by taking Absolute value, normalizing the audio, subtracting the normalized arrays, and taking the sum of absolute differences. The segment that shows the least error is considered the candidate for the match.

The matching performance proved to be dependent on the complexity on the music. Matching simple music, such as single-note pieces, was successful. However, more complex pieces, such as Chopin's Ballade 4, was not successful, that is, the algorithm couldn't produce low error value in any of the music in the database. I suspect that it has to do with having "too many notes," i.e., mismatches in the higher harmonics added up to significant amount of errors, which swamps the calculations.

# Introduction & Background Research

There are moments when one hears a beautiful piece of music but doesn't know the name of it. I wish there were a way to find out. Indeed, there is an app called Shazam which somewhat achieves this goal. However, Shazam does not match music. Instead, it identifies a particular recording of a particular music [1]. In fact, it cannot match the same music played by the same musician if they are two different recordings [1]. This is because Shazam isn't comparing frequencies but, instead, it is going by the loudness variation of the music in time [2].

But this isn't how humans perceive music. People hear pitches, and harmonies. Our algorithm tries to humanize the music recognition by emphasizing these two factors by using spectrogram. When somebody whistles or hums a melody, even if the rhythm and notes are different, I hope to find the correct match for the piece.

According to Wolfram Documentation, Spectrogram plots the magnitude of the short-time Fourier transform (STFT), computed as a discrete Fourier transform (DFT) of partitions of list [3]. In simple terms, Spectrogram is a visual way of representing the signal, and strength, or "loudness," of a signal over time at various frequencies present in a particular waveform [4]. By visualizing Spectrogram in 3D as shown in Figure 1, I can see the fundamental frequency in the front and the harmonics building up behind it. The log scale was used to make the visuals clearer. For spectrograms that depict audios, its fundamental frequency is the clearest marker in the graph, always the bottom most one. In Spectrograms, the darker the line is, the louder its volume is. Figure 2 illustrates the following. The combination of all the harmonics in the spectrogram results in the timber of the sound.

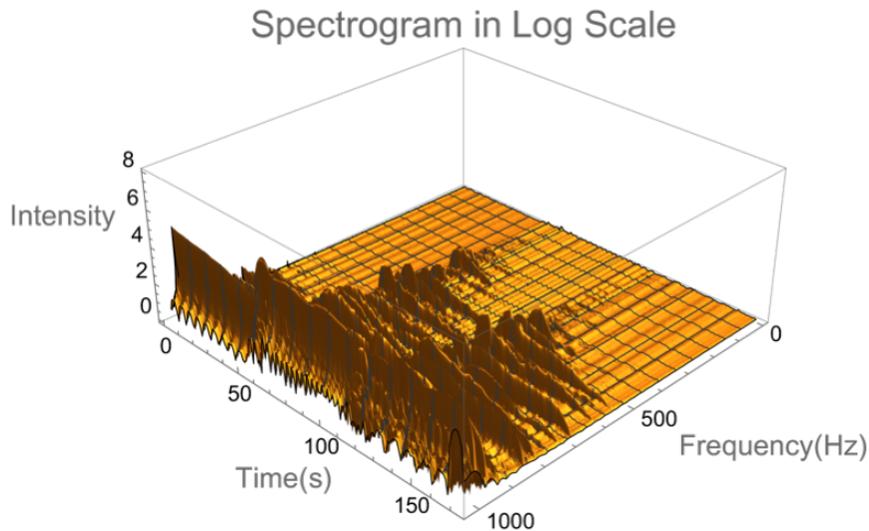

*Figure 1 Spectrogram in Log Scale*

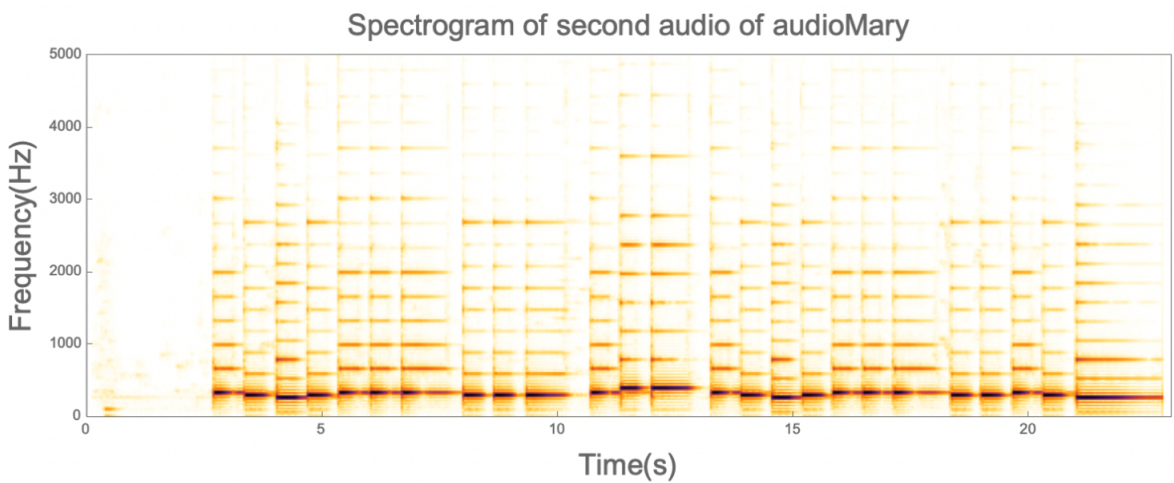

*Figure 2 Spectrogram of second audio of audioMary*

The `SpectrogramArray` returns the data that went into the construction of the `SpectrogramArray`, allowing us to use the data from our analysis. Since the Fourier Transform used in the `SpectrogramArray` returns complex numbers, the absolute value of it was used [5].

Another key part of the algorithm was the `Partition` command. `Partition` cuts lists into smaller portions. In Mathematica, by specifying an offset value, it was possible to cut the list with overlaps [6]. This was very useful for this code because there were more parts of the long music to compare the sample music to, which translated to the result being more accurate.

# Methods & Materials

Materials:

1. Sample pieces of music from YouTube. For testing, three sample music performances were used: Mary Had a Little Lamb, Cello scales, and Chopin's Ballade 4. Lengths do not have to be the same.

2. Wolfram Mathematica version 12.3

Methods:

### 1. Baseline for writing the algorithm

First, I stored the files in the folders and read the names of the audio files. Also, I get all the files to be used as databases, i.e., music to be compared against, and store them in a folder. The next step is to turn every music into spectrogram arrays. When I am matching the unknown piece, find the Spectrogram Array of it, then match it against every piece in the database, one at a time, keeping track of the error. The one with the least error will be declared the match if the error value is reasonable.

### 2. Spectrograms & `mySpectro` function

As stated in the introduction, spectrograms are a visual way of representing the signal strength, or "loudness," of a signal over time at various frequencies present in a particular waveform. Mathematica has a great built-in function, and Figure 3 is a spectrogram of one of the audios used.

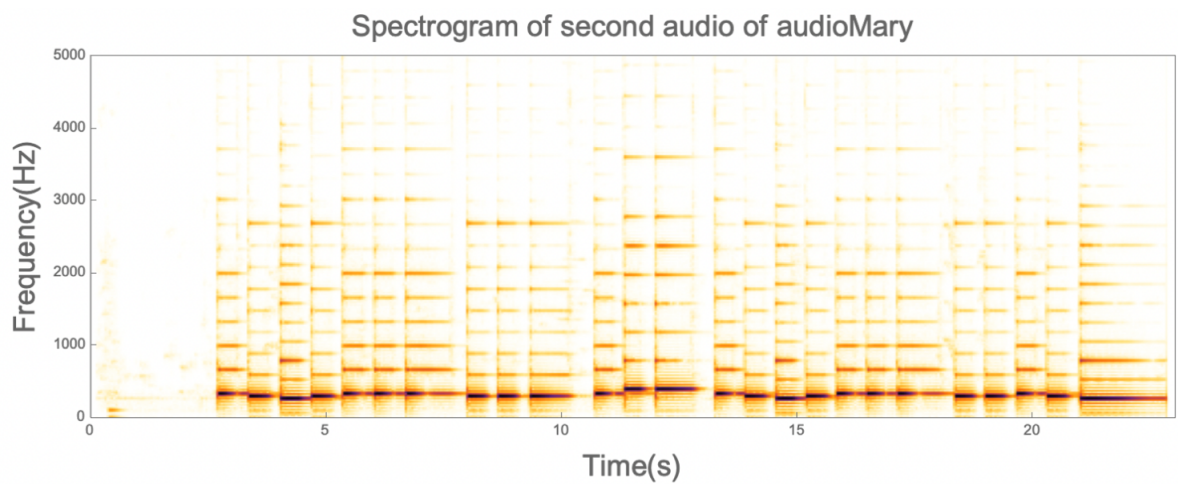

*Figure 3 Spectrogram of second Audio of audioMary*

Anthony Lee

It is clear the darker notes towards the bottom, which represent the fundamental frequencies, show the pattern and progression of the notes of *Mary Had a Little Lamb*. However, because `Spectrogram` is a plot, it is very difficult to extract any quantitative information from it. Therefore, I used `SpectrogramArray` instead, which gives the actual data. Figure 4 shows the `mySpectro` function which gives us the spectrogram array data from an arbitrary segment within a music specified by its parameters.

```
mySpectro[audio_, {start_, end_}] :=
 Module[{audio2, duration}, duration = Duration[audio];
  If[{start, end} = {0, 0}, audio2 = audio,
   If[start = 0, audio2 = AudioTrim[audio, end],
    If[end = 0,
     audio2 = AudioTrim[audio, {start, duration}],
     audio2 = AudioTrim[audio, {start, end}]

    ];
   ];
  ];
 Abs[SpectrogramArray[AudioChannelMix[audio2],
    4096, 512][All, -1024 ;; -1]]
 ]
```

*Figure 4 mySpectro function*

The `AudioChannelMix` was used to convert stereo sound into mono so that audio could be match one to one. The two channels were averaged to produce the center-panned stereo audio object. In the function, the length of the partitions (Fourier Transform time window) was set to 4096 samples for all the audio and the offset value was set to 512 samples.

```
ListPlot3D[testSegmentArray,
 AxesLabel → {Style["Frequency(Hz)", FontSize → 13],
   Style["Time(s)", FontSize → 13], Style["Intensity", FontSize → 13]},
 PlotLabel → Style["Spectrogram in Linear Scale", FontSize → 17],
 PlotRange → {0, 500}]
```

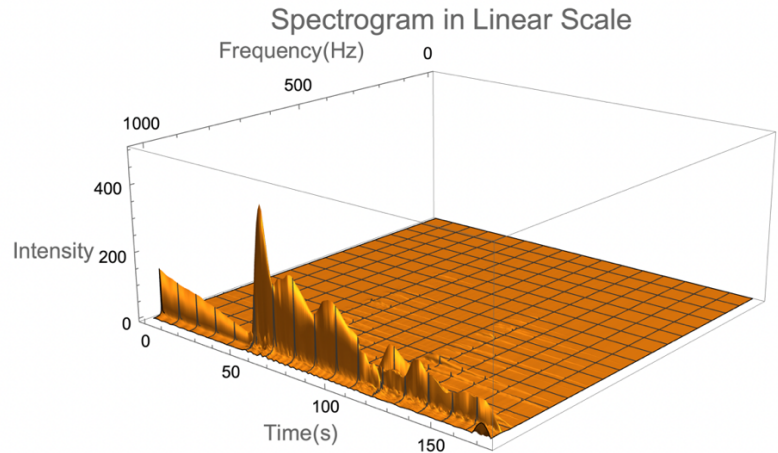

*Figure 5 Spectrogram in Linear Scale #1*

```
ListPlot3D[testSegmentArray2,
 AxesLabel → {Style["Frequency(Hz)", FontSize → 13],
   Style["Time(s)", FontSize → 13], Style["Intensity", FontSize → 13]},
 PlotLabel → Style["Spectrogram in Linear Scale", FontSize → 17],
 PlotRange → {0, 500}]
```

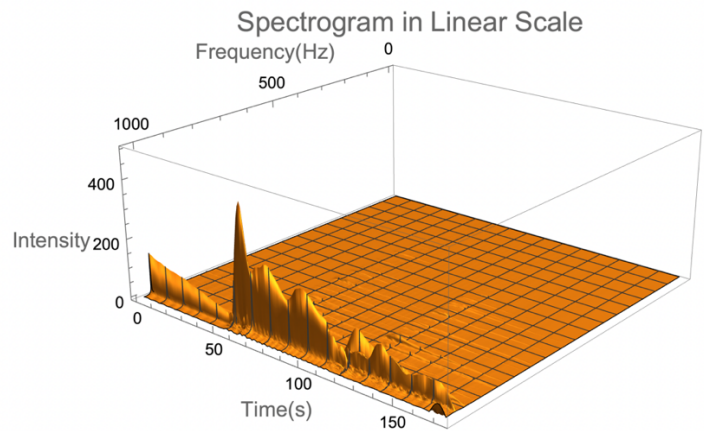

*Figure 6 Spectrogram in Linear Scale #2*

After seeing the `ListPlot3D` of both in a linear scale, I confirmed that `mySpectro` works.

### 3. Partitioning `SpectrogramArrays`

Now that `mySpectro` is complete, the next step was to partition the music into the length of the small music segment. Figure 8 shows what Partition looks when Figure 7, the spectrogram, is partitioned.

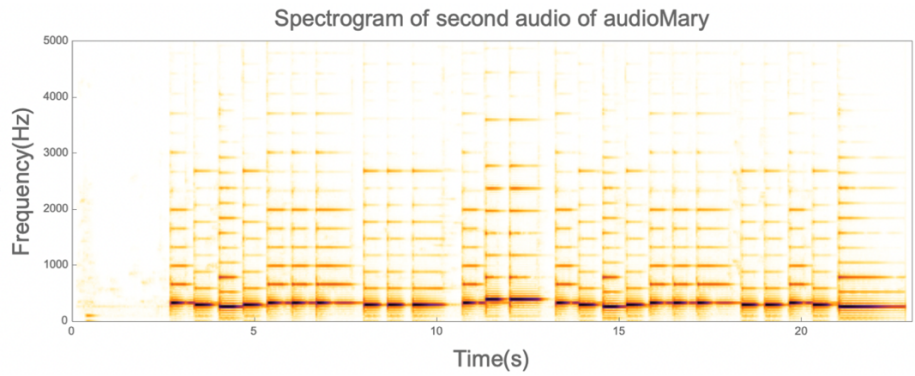

*Figure 7 Spectrogram*

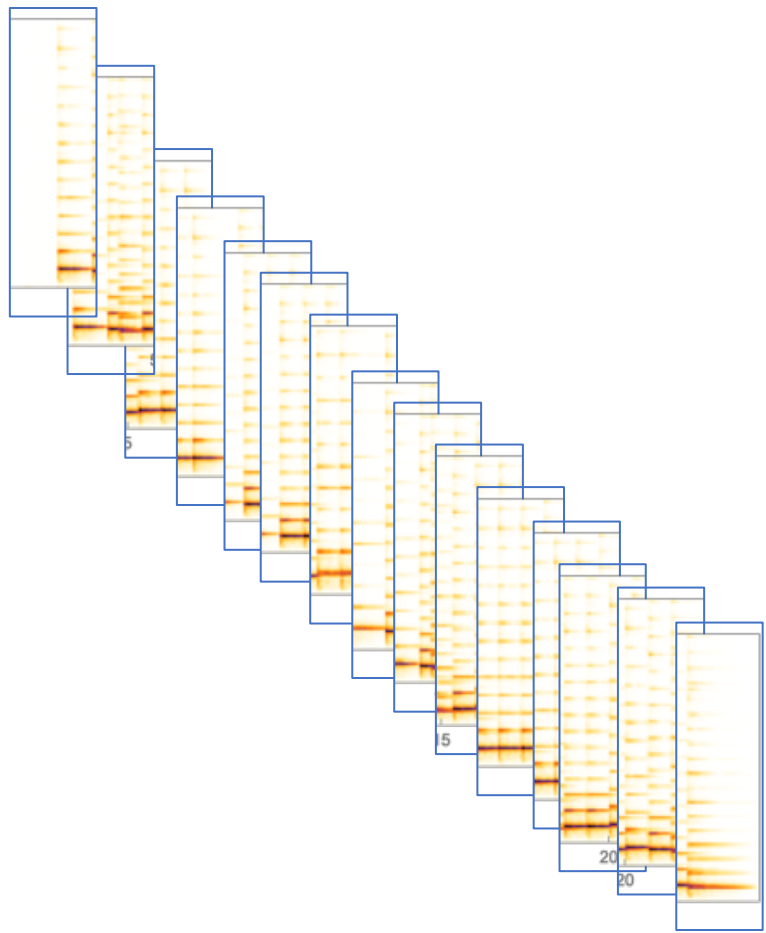

*Figure 8 Illustration of Partition*

I partitioned the songs in the database to match the unknown size with the offset value of 10 sample size. The offset value means that partitioning is done with the overlap of 10 samples.

### 4. Audio Normalization

The whole idea of Audio Normalization is to avoid Shazam's way of matching the volume variations. Thus, I wanted the matching to be immune to the loudness of particular performances or recordings by normalizing the music volume. If one performance is recorded at a higher volume than the other, even if the pieces are played perfectly matching, there will be errors when the spectrograms are subtracted because there will be height differences in them. Audio Normalization ensures that the maximum volume of each music segment is the same. As loud music has higher values (manifested as taller heights in 3D, darker colors in 2D) in spectrogram, normalization corrects for this variation reducing false error.

*Figure 9 Audio Normalizing*

By dividing each element by the Max volume of each element, the values in the spectrum array becomes a volume ratio to the max, rather than absolute volume. This makes the comparison more apple to apple when comparing recordings of different volumes. The +1 added to the max value seen in the code was added in the denominator to prevent dividing by 0 when the max is zero. This could happen if the music was silent. The +1 doesn't affect the value of the spectrum because they are in the thousands and +1 was added equally to both the database and the unknown segment. Therefore, if the music is truly identical, their error should be zero after their point-by-point subtraction.

### 5. Subtracting `SpectrogramArrays`

Now, I subtracted the spectrogram arrays of the unknown from the ones in the database.

```
errors = Table[{i, Total[Abs[testSegmentArrayNorm - partitionedNorm[[i]]], 2]}, {i, 1, Length[partitioned]}]
```

{{1, 603.785}, {2, 677.229}, {3, 747.634}, {4, 814.787}, {5, 887.518}, {6, 958.259}, {7, 447.19},
{8, 403.691}, {9, 426.913}, {10, 433.516}, {11, 437.583}, {12, 433.351}, {13, 415.078}, {14, 394.436},
{15, 362.583}, {16, 345.025}, {17, 326.568}, {18, 298.917}, {19, 348.408}, {20, 418.859}, {21, 464.104},
{22, 488.003}, {23, 511.827}, {24, 513.547}, {25, 514.816}, {26, 610.601}, {27, 603.265}, {28, 619.948},
{29, 632.625}, {30, 637.633}, {31, 563.623}, {32, 582.884}, {33, 574.325}, {34, 537.294}, {35, 498.962},
{36, 481.452}, {37, 482.791}, {38, 483.959}, {39, 461.523}, {40, 432.885}, {41, 360.716}, {42, 394.428},
{43, 422.601}, {44, 418.705}, {45, 415.646}, {46, 389.964}, {47, 357.42}, {48, 398.864}, {49, 392.505},
{50, 385.1}, {51, 366.403}, {52, 320.298}, {53, 328.749}, {54, 375.458}, {55, 421.922}, {56, 451.567},
{57, 476.149}, {58, 493.58}, {59, 494.481}, {60, 540.737}, {61, 539.746}, {62, 524.567}, {63, 504.011},

*Figure 10 Table obtained by subtracting spectrogram arrays*

Figure 10 illustrates the process of **testSegmentArrayNorm** and elements of **partitionedNorm** being compared. The goal was to find the element in **partitionedNorm** with the least error, which meant that the segment was found in the large music. To do this, **Total** was calculated which sums up all the Absolute values of the differences because error is defined to be the sum. (There were more data points, but was cut off because of space)

```
errors2 = {rescale[#[[1]]], #[[2]]} & /@ errors;
```

*Figure 11 Rescaling x-axis*

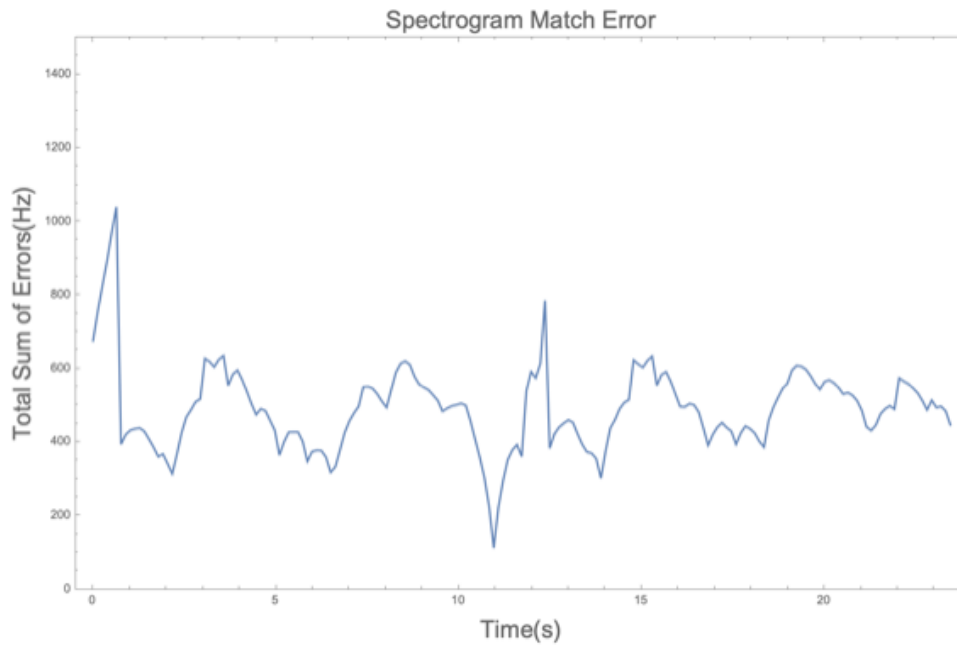

*Figure 12 ListLinePlot of errors2*

Figure 12 shows the **ListLinePlot** of errors2. Clearly, the least error happened around at 11th seconds. Going back to **testSegmentArray**, 11 is indeed the center of the intervals of 10 and 12, which confirms the correct match.

To ensure that the match wasn't a coincidence, another test with the interval of 10 and 14 seconds of the music was done.

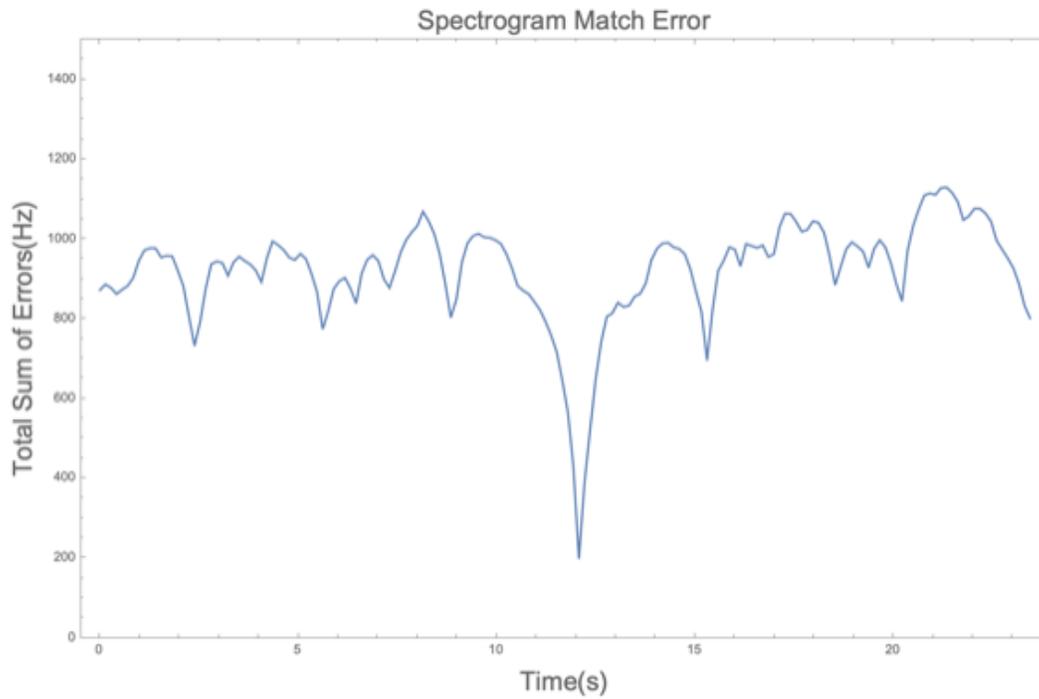

*Figure 13 ListLinePlot of another test*

Figure 13 shows that the least error occurs at 12 seconds, which is the center of the interval between 10 and 14 seconds, which again confirms the result.

### 6. Building the `myErrorFinal` function

Here is the myErrorFinal function, which makes the matching much more convenient.

```
myErrorFinal[audioFull_ , audioSegment_ , stepSize_ : 1] :=
 Module[{testAudio, duration, testSegmentAudio, testArray, testSegmentArray,
    testSegmentArrayNorm, tpTimeSegment, partitioned, partitionedNorm,
    minMax, rescale, errors},
  testAudio = audioFull;
  duration = QuantityMagnitude[Duration[testAudio]];
  testSegmentAudio = audioSegment;
  testArray = mySpectro[testAudio, {0, 0}];
   testSegmentArray = mySpectro[testSegmentAudio, {0, 0}];
   testSegmentArrayNorm = testSegmentArray / (Max[testSegmentArray] + 1);
  tpTimeSegment = Length[testSegmentArray];
  partitioned = Partition[testArray, tpTimeSegment, stepSize];
  partitionedNorm = # / (Max[#] + 1) & /@ partitioned;
  minMax = {1, Length[partitioned]};
  rescale[index_] := Rescale[index, minMax, {0, duration}];
  errors =
     Table[{rescale[i], Total[Abs[testSegmentArrayNorm - partitionedNorm[[i]]], 2]},

     {i, 1, Length[partitioned]}];
 Image[ListLinePlot[errors, Frame → True,
    FrameLabel → {Style["Time(s)", FontSize → 20],
      Style["Total Sum of Differences of Frequency(Hz)", FontSize → 20]},
    PlotLabel → Style["Total Sum of Differences of Frequency(Hz) vs Time(s)",
      FontSize → 20], ImageSize → 800, PlotRange → {0, 2000}]]
 ]
```

*Figure 14 myErrorFinal function*

Figure 14 shows the `myErrorFinal` function.

Here is a brief step-by-step description of how the function:

- o  Function trims the music based on its duration using `mySpectro` and gets the spectrogram arrays
- o  The `Partition` command partitions the large spectrogram array into smaller portions that match the length of the sample, which is two seconds.
- o  The music volume was normalized when the music was completely silent.

- o  Partitioned spectrogram arrays were subtracted from the short spectrogram array and the `ListPlot` was drawn to identify the seconds with the least errors. The least errors were represented with the lowest `Total` value of the absolute values of the differences.

7. **Testing the `myErrorFinal` function.**

   `myErrorFinal[audioMary〚3〛, AudioTrim[audioMary〚3〛, {6, 8}], 1]`

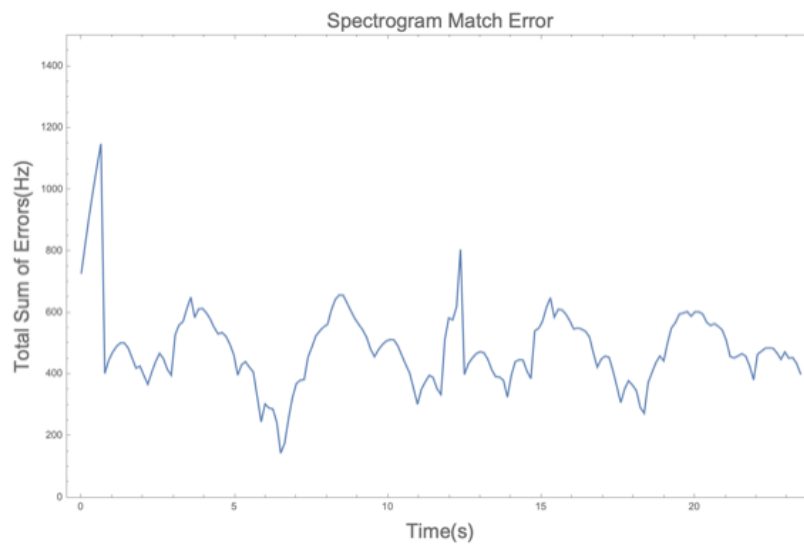

*Figure 15 Testing between {6, 8} seconds*

Figure 15 shows testing that was done with the segment of `{6,8}` seconds of the third audio of `audioMary`. Even though the `ListLinePlot` shows the least number is somewhere around 6.7, it is very close to 7, which is the center of the interval of 6 and 8 seconds and can conclude that this was a successful test.

# Results & Discussion

### 1. Successful Cases of *Mary Had a Little Lamb*

The algorithm was tested to see if it could identify the matching segment for identical piano pieces from the same recording. As shown in the Methods & Materials section, the clear low error values confirmed that identification worked. Figure 16 shows a clear low error at 12 seconds, which is indeed the center of the interval between 10 & 14.

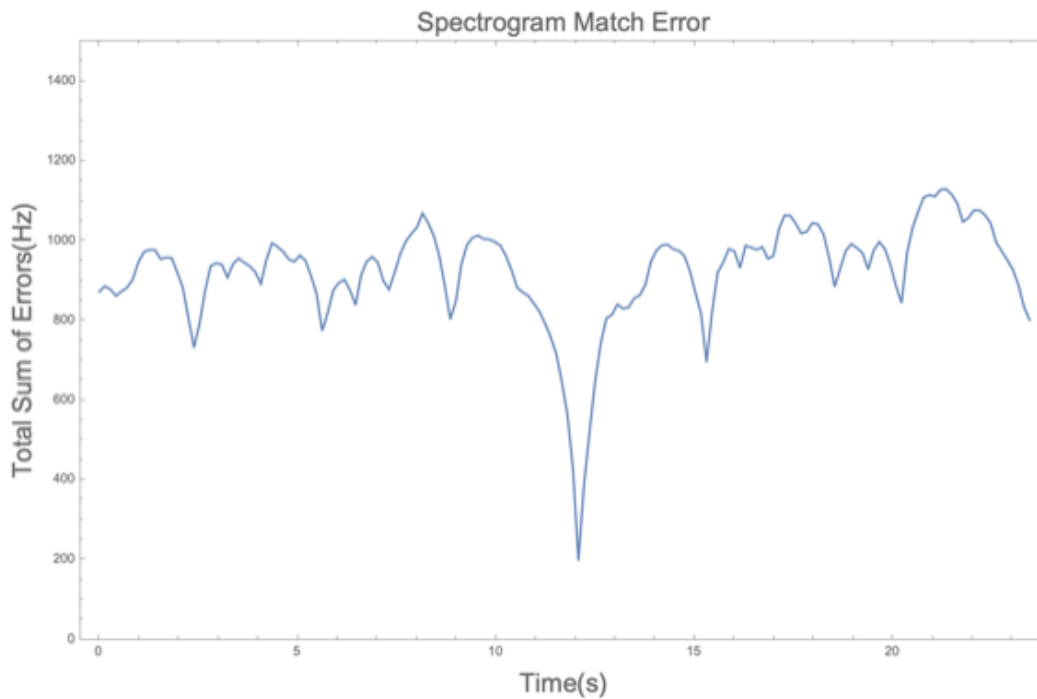

*Figure 16 Errors between 10&14 seconds*

### 2. Some Unorthodox Cases in matching Mary Had a Little Lamb

Not all the results for `audioMary` were as expected. The plot showed that *Mary Had a Little Lamb* had many parts of repetition. It would be useful in studying musicology when studying how many times the theme was occurring in the music.

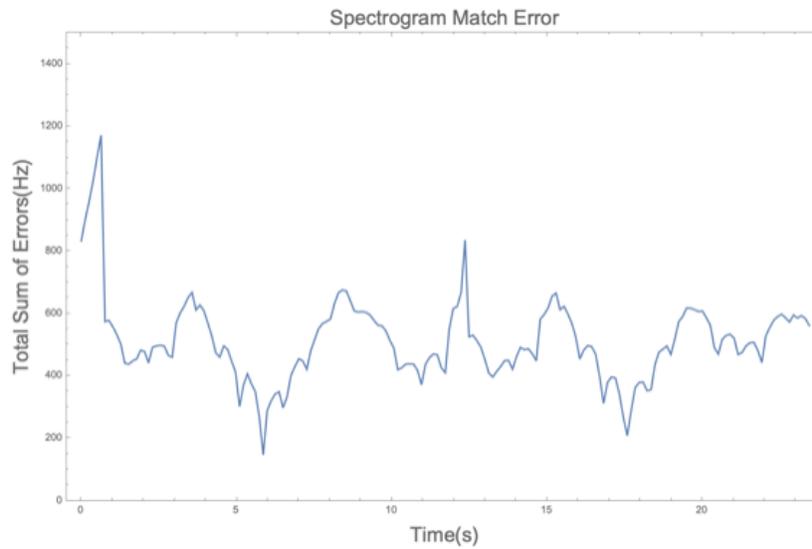

*Figure 17 Errors between 16 & 18 seconds*

Figure 17 shows the two low errors around 6 seconds and 17 seconds. This happens because the notes repeated. Indeed, I can confirm this by listening to the music. However, because our original intent in this research was matching the pieces, and the location of the match, so having multiple low error locations is beneficial for verifying that the match was correct. Also, the symmetry that happens at 6 seconds and at 17 seconds are very interesting too.

### 3. Matching Mary Had a Little Lamb played by two different recordings.

This was tests to show that even when the same piece is played by two different players, the algorithm still can identify the matching segment with the least error.

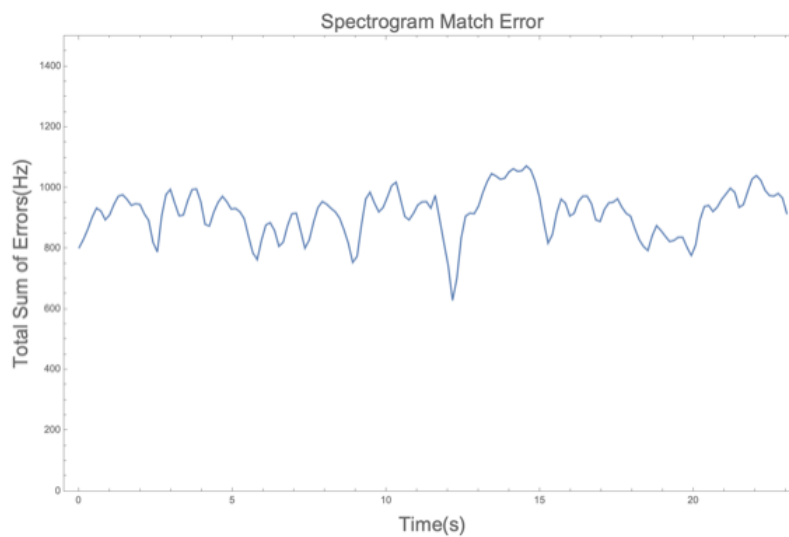

*Figure 18 Matching using audios played by different people.*

Figure 18 shows the low error at 12 seconds, which is the center of 10 and 14. Thus, the algorithm can find the matching segment in the case when the same music is played by two different people. However, compared to tests when the same player plays identical pieces, the errors are greater. Nonetheless, the algorithm found the correct match. This type of matching is one that Shazam cannot do. More tests were conducted to verify if it indeed worked, and the test results showed clear low errors unless the segment was repeated at another time in the music.

### 4. Cello Music Matching

The tests with cello music were conducted to see if the algorithm could be expanded into different instruments. The cello music was a cello playing scales of five notes with breaks in the middle, as seen in Figure 19.

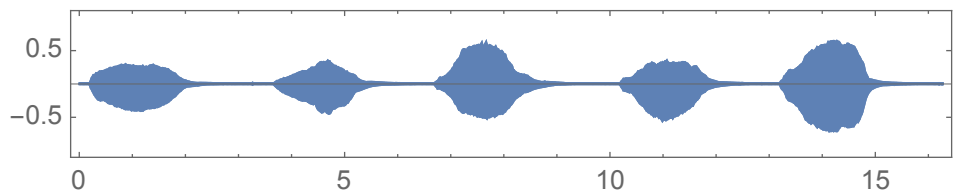

*Figure 19 AudioPlot of cello audio*

The algorithm was able to find the correct match as Figure 20 illustrates in the interval of 10 & 12. The matching was found at the 11[th] second.

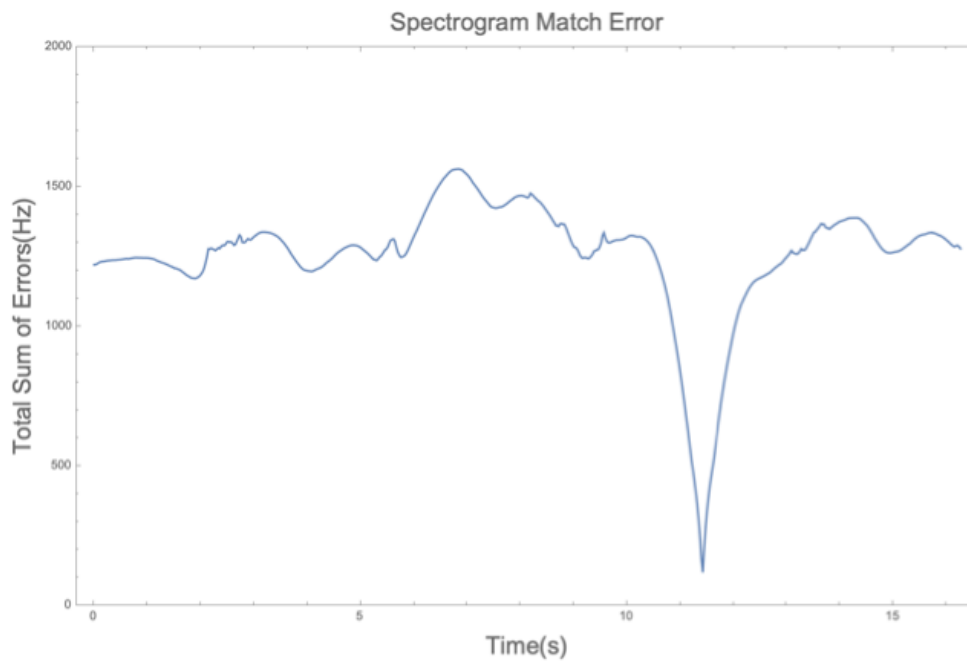

*Figure 20 Cello matching between 10 and 12 seconds*

### 5. Matching a more complex piano piece with the same recording.

Now that matching with single note pieces was successful, more complex pieces were tested. The example used was Chopin's Ballade 4, a very rich and complex piano piece. A 30 second segment trimmed and out of the 30 second segment another one second segment was trimmed. Figure 21 shows the low error value at 5.5, which shows that the matching was successful since the example music was chosen between 5&6 seconds. Considering how this algorithm works, complex pieces are harder to match.

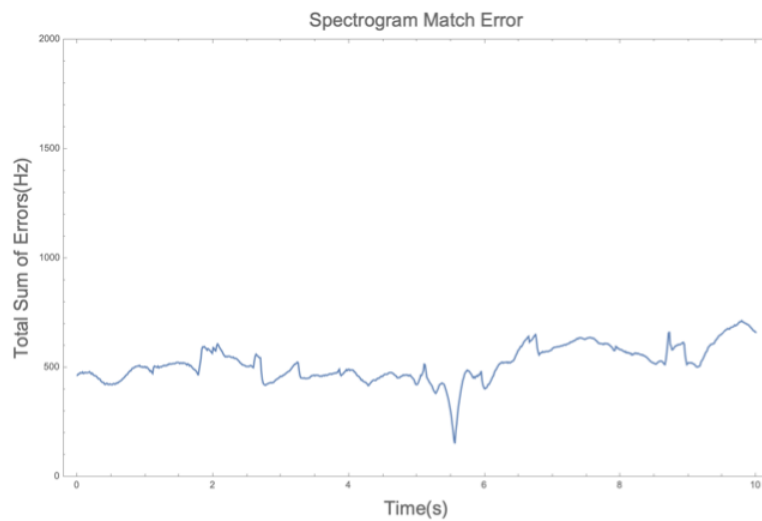

*Figure 21 Chopin's Ballade 4 matching between 5 & 6 seconds*

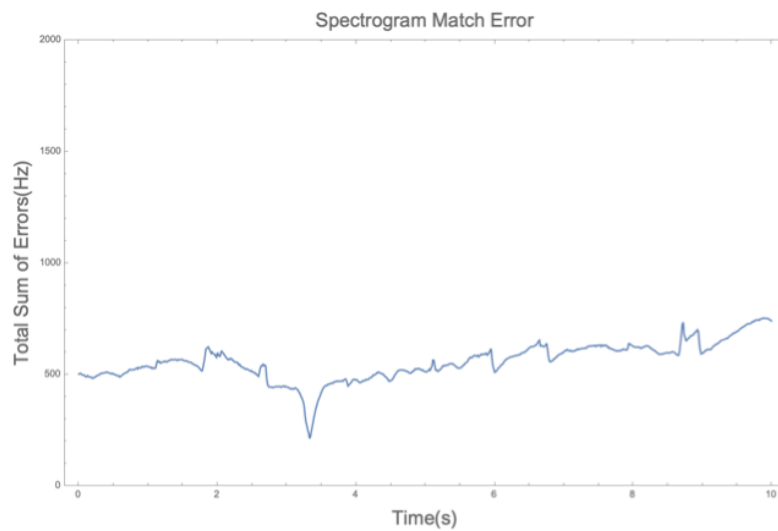

*Figure 22 Chopin matching between 3&4 seconds*

Figure 22 shows the matching between 3&4 seconds, and the low error value occurs at 3.5 seconds, which is very accurate.

**6. Matching with a more complex piano piece played by different players.**

This test was done to see if the algorithm could match a complex piece played by different players. The test was conducted with Chopin's Ballade 4. Unfortunately, the match was unsuccessful as there wasn't a certain low error value that was evident. Also, at 8.5 seconds, which was where the matching should have been, there isn't a clear low error value. Figure 23 illustrates this occurrence.

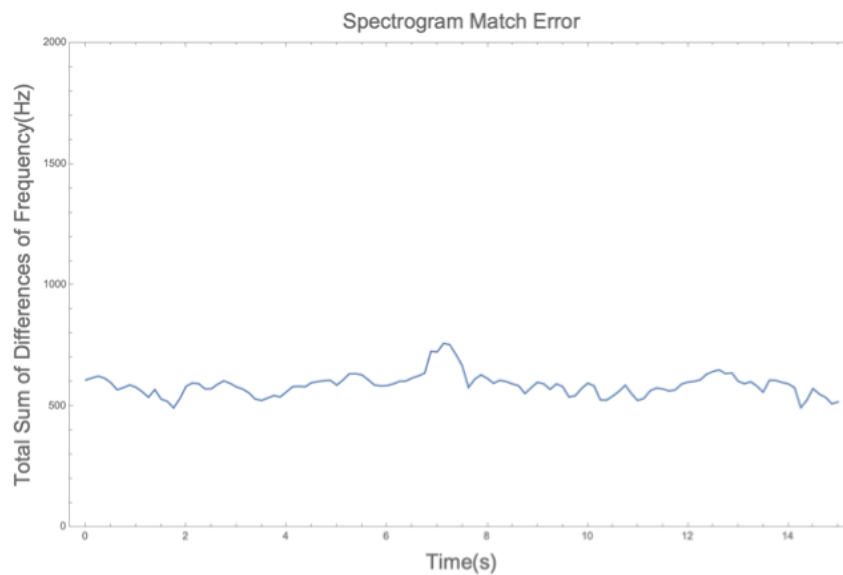

*Figure 23 Chopin Ballade 4 matching played by different players*

**7. Difference between Normalized and not Normalized Audio.**

The fourth step in the Methods & Materials section was Audio Normalization, which made the maximum volume for each segment of the music that I am comparing to be equal to 1. The side effect of our intensity affecting our error was minimized. However, it is interesting how the plot of errors still was able to match the low error values. Figure 24 is the errors between 15 & 17 seconds before normalization and Figure 25 is the errors after normalization.

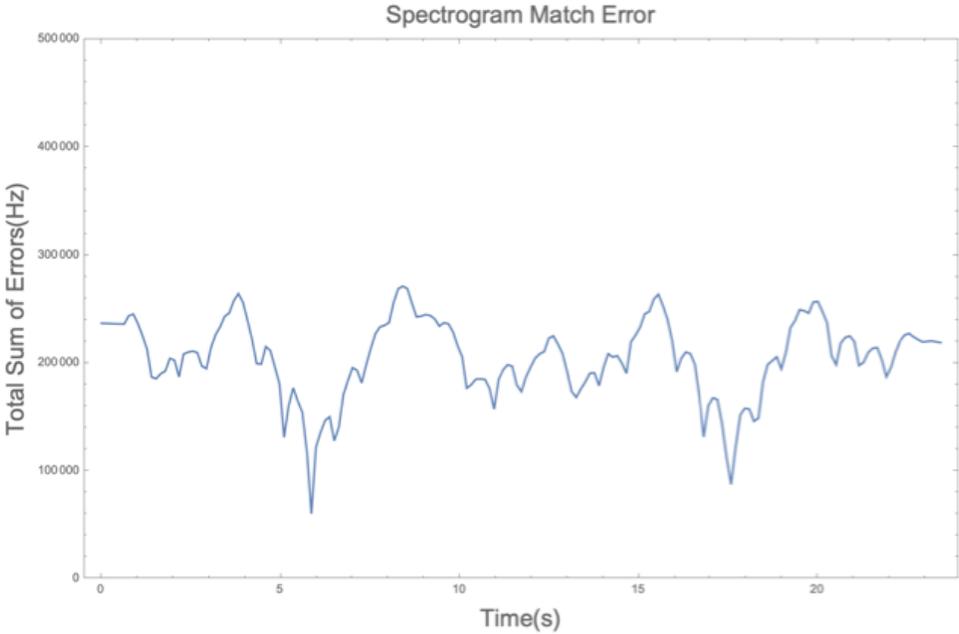

*Figure 24 Errors before Normalization*

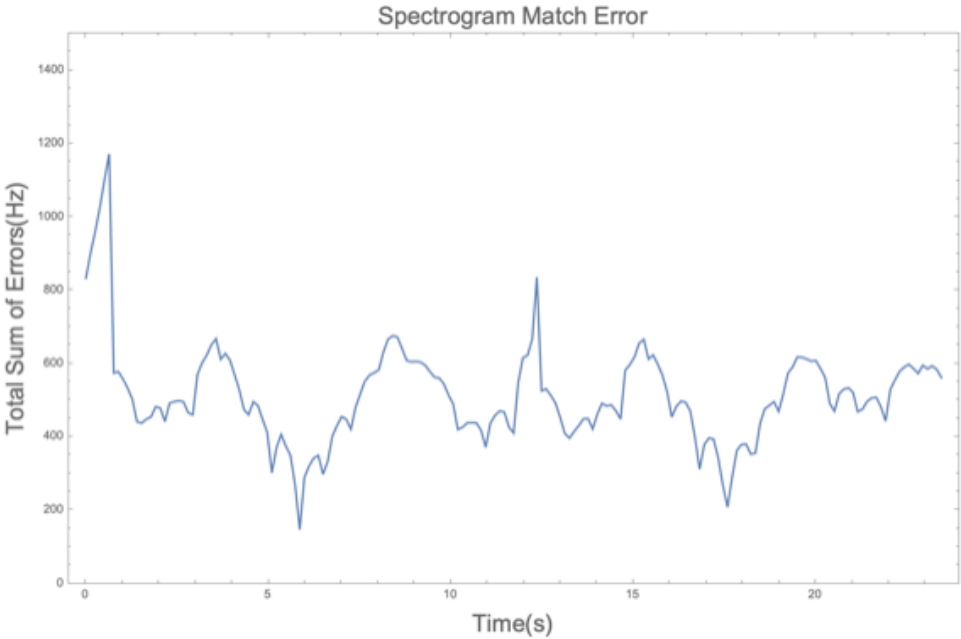

*Figure 25 Errors after Normalization*

The main difference between the figures are the error values. The errors of Figure 25 are around 300, while the y values for Figure 24 are around 40,000, which shows how crucial normalization is.

**8. Example of complete mismatch: errors between ocean sound and piano sound.**

The goal of this exploration was to see what the errors graph looked like when totally unrelated audios were put in. One of Mary Had a Little Lamb was chosen and a one second sound of ocean was chosen. As shown in Figure 26, the graph is almost a straight line without any low error region as expected.

`myErrorFinal[audioMary〚3〛, AudioTrim[ocean, {3, 4}], 10]`

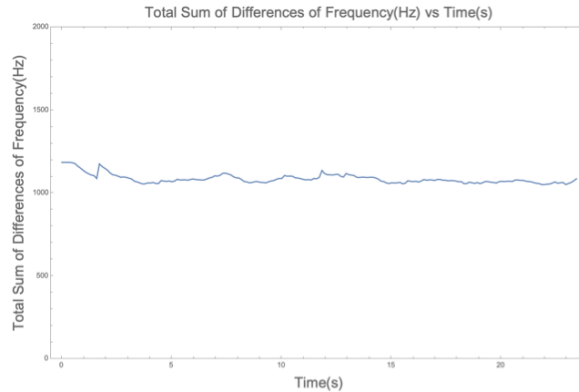

*Figure 26 Errors or Ocean & AudioMary*

## Conclusion & Future Research

Humans can recognize an orchestra piece by humming or whistling the melody line. This research attempted to get closer to how humans identify music. Based on experimental results when music with only single notes is played matching was very successful. However, once single note music was played by different recordings, the success rate decreased. When a portion of a complex pieces was compared to the whole, the algorithm found where the portion came from in the timeline. However, when the portion was compared to different recordings of the same music, matching wasn't successful.

For future research, the goal is to improve this algorithm by making sure it can match complex pieces played by different recordings.